\documentclass[twocolumn,secnumarabic,amssymb,nobibnotes,aps,prl,noeprint]{revtex4-1}
\usepackage[utf8]{inputenc}
\usepackage{blindtext}
\usepackage{chemformula} 
\usepackage[T1]{fontenc}
\usepackage{graphicx}
\usepackage{natbib}
\usepackage{textgreek}
\usepackage{siunitx}
\usepackage[utf8]{inputenc}
\usepackage{amsmath}
\usepackage{xcolor}

\newcommand{\KR}{KR }

\begin{document} 

\author{Lukas Powalla$^{1,2}$*, Jonas Kiemle$^{3,4}$*, Elio J. König$^{1}$, Andreas P. Schnyder$^{1}$, Johannes Knolle$^{4,5,6}$, Klaus Kern$^{1,2}$, Alexander Holleitner$^{3,4}$, Christoph Kastl$^{3,4}$ and Marko Burghard$^{1}$}
\affiliation{
1 Max-Planck-Institut für Festkörperforschung, Heisenbergstrasse 1, D-70569 Stuttgart, Germany.\\
2 Institut de Physique, Ecole Polytechnique Fédérale de Lausanne, CH-1015 Lausanne, Switzerland.\\
3 Walter Schottky Institut and Physics Department, Technical University of Munich, Am Coulombwall 4a, D-85748 Garching, Germany.\\
4 MCQST, Schellingstrasse 4, D-80799 München, Germany.\\
5 Department of Physics TQM, Technical University of Munich, James-Frank-Strasse 1, D-85748 Garching, Germany.\\
6 Faculty of Natural Sciences, Department of Physics, Imperial College London, London SW7 2AZ, UK.}
\date{\today}                                                                                                                                                                                                                                                                                                                                                                     


\title{Berry curvature-induced local spin polarisation in gated graphene/WTe$_2$ heterostructures}
\maketitle


\textbf{Full experimental control of local spin-charge interconversion is of primary interest for spintronics \cite{ahn20202d, avsar2020colloquium}. Heterostructures combining graphene with a strongly spin-orbit coupled two-dimensional (2D) material enable such functionality by design \cite{gmitra2015graphene,NovoselovCastroNeto2016, vzutic2019, Island2019}. Electric spin valve experiments have provided so far global information on such devices \cite{ghiasi2017large, safeer2019large,ghiasi2019charge,safeer2019room, khokhriakov2020gate, zhao2020observation, hoque2020charge,li2020gate,kovacs2020electrically, benitez2020tunable}, while leaving the local interplay between symmetry breaking, charge flow across the heterointerface and aspects of topology unexplored. Here, we utilize magneto-optical Kerr microscopy to resolve the gate-tunable, local current-induced spin polarisation in  graphene/WTe$_2$ van der Waals (vdW) heterostructures. It turns out that even for a nominal in-plane transport, substantial out-of-plane spin accumulation is induced by a corresponding out-of-plane current flow. We develop a theoretical model which explains the gate- and bias-dependent onset and spatial distribution of the massive Kerr signal on the basis of interlayer tunnelling, along with the reduced point group symmetry and inherent Berry curvature of the heterostructure. Our findings unravel the potential of 2D heterostructure engineering for harnessing topological phenomena for spintronics, and constitute an important further step toward electrical spin control on the nanoscale.}

2D quantum materials are of immense interest as components of novel quantum electronic, spintronic and optical devices~\cite{NovoselovCastroNeto2016,ahn20202d, avsar2020colloquium}. Combining them into vdW heterostructures allows for tailoring their electronic properties through control of the interfacial wave functions' symmetry, spin-orbit coupling (SOC) or proximity exchange, thereby imparting for instance superconductivity or magnetism \cite{gmitra2015graphene,vzutic2019}. Complementing interface design with external control via gate fields \cite{Island2019}, the electronic properties can be manipulated on-demand, enabling for example switchable topology or spintronic functionality. In particular, graphene as a spin transport channel combined with a strong SOC 2D material as spin generator enables a gate-tunable spin-charge interconversion \cite{ahn20202d, avsar2020colloquium}, as has been demonstrated for vdW heterostructures incorporating a 3D topological insulator \cite{khokhriakov2020gate}, a trivial or topological 2D semimetal \cite{zhao2020observation,safeer2019large, hoque2020charge}, or a 2D semiconductor \cite{li2020gate, ghiasi2019charge,safeer2019room,ghiasi2017large,kovacs2020electrically}. The electrically detected spin-signal can even persist up to room temperature, which is highly desirable for applications \cite{khokhriakov2020gate, zhao2020observation, hoque2020charge, li2020gate,ghiasi2019charge, benitez2020tunable,kovacs2020electrically}. 

To elucidate more intricate properties of the heterostructures, in particular those associated with topology, we use Kerr rotation (KR) microscopy to probe the local, gate-dependent spin-charge interconversion. In general, a spin accumulation $\langle S_z \rangle$ induces a 2D anomalous Hall conductivity $\sigma_H$, which by Maxwell's equations implies a polar Kerr angle $\theta_{K}$. At the same time, the local anomalous Hall conductivity is given by the integral of the Berry curvature over locally filled states, and therefore  vanishes in equilibrium in the presence of time reversal invariance. However, non-centrosymmetric materials allow for a finite Berry curvature dipole $D^{(\Omega)}_{ji} $~\cite{SodemannFu2015} and  $\sigma_H(\mathbf x)  \propto  D^{(\Omega)}_{jz} j_j(\mathbf x)$ can be induced by a local diffusion current $ j_j(\mathbf x)$ inside the topological metal, leading to 
\begin{equation} \label{eq:RelSpinTopo}
    \langle S_z \rangle(\mathbf x) \propto \theta_{K}(\mathbf x) \propto D^{(\Omega)}_{jz} j_j(\mathbf x).
    \end{equation}

\begin{figure*}[t]
\includegraphics[scale=0.98]{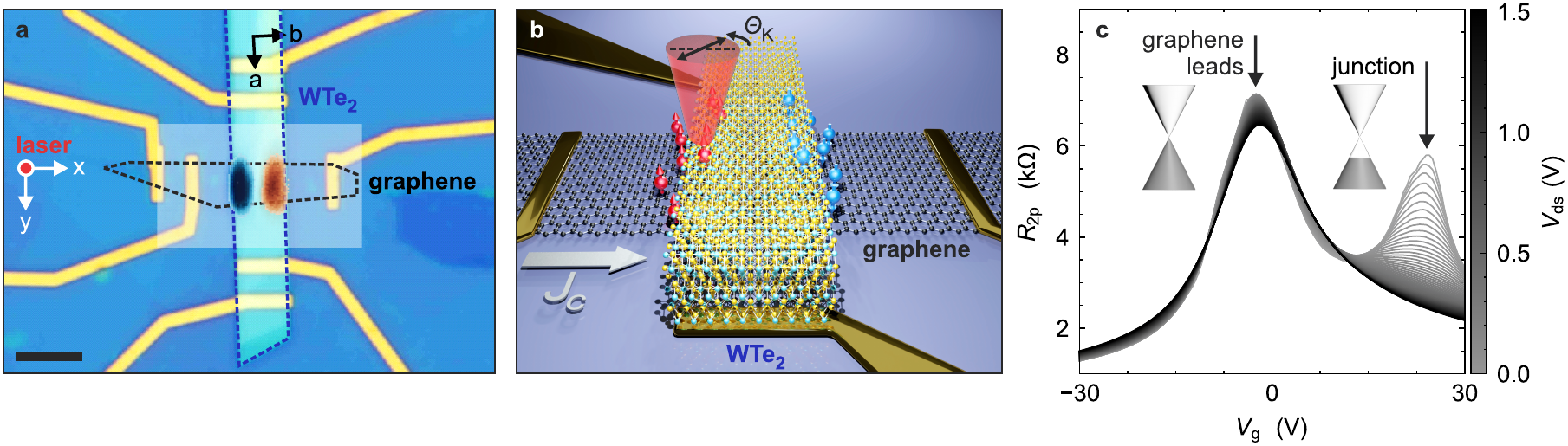}
\caption{\textbf{Optical KR microscopy and graphene transfer curves}. (a) Optical microscope image of a heterostructure comprised of graphene (black dashed line), WTe$_2$ (blue dashed line) and hBN capping on a back-gated \ch{Si/SiO2} substrate. The WTe$_2$ crystallographic axes are indicated. The overlay shows the current-induced KR signal at the junction using a colour code as in Fig.~\ref{fig: 2}. Scale bar, \SI{5}{\micro\meter}. The probed area is marked by the shaded rectangle. (b) An oscillating current at frequency $\omega$ (horizontal arrow) is applied along the graphene stripe. At the heterojunction, the current-induced out-of-plane spin component (red and blue arrows) is locally read-out by a polar KR measurement at frequency $\omega$. The effect of Joule heating is detectable at twice the oscillation frequency $2\omega$. (c) Transfer curves of the graphene stripe within the device in panel (a), measured in 2-probe configuration at $T_\text{bath} = \SI{4.2}{\kelvin}$ using an ac voltage that increases stepwise up to $\SI{1.5}{\volt}$. The two Dirac points reveal different doping levels (see inset) for the bare graphene leads (charge neutral) and the proximitized graphene ($p$-doped).}
\label{fig: 1}
\end{figure*}

This establishes a relationship between Kerr microscopy, spintronics, topological band theory, and non-reciprocal transport coefficients~\cite{TokuraNagaosa2018,stamm2017magneto,LeeShan2017}, which also determine the non-linear anomalous Hall effect~\cite{KangMak2019,MaJarilloHerrero2019, culcer2020transport} and photocurrents in topological metals~\cite{XuJarilloHererro2018,Wang2019, Kiemle2020b}. While the current-induced anomalous Hall response in the bulk of either graphene or WTe$_2$ vanishes due to crystalline symmetries,
we here exploit the reduced interface symmetry of graphene/WTe$_2$ heterostructures to achieve a controllable net spin-charge interconversion.


The investigated heterostructure consists of an approximately 20 nm thick  WTe$_2$ ribbon interfaced to a single-layer graphene stripe beneath, such that the two long axes form an angle close to 90° (Fig. \ref{fig: 1}a). The long (short) axis of the WTe$_2$ crystal corresponds to its $a$-axis ($b$-axis) (Supplementary Fig. S1). The $c$-axis points out-of-plane. The overlay in Fig. 1a shows a current-induced KR map, indicating a substantial out-of-plane spin generation locally within the junction. The red and blue areas denote KR signals, i.e. spin polarisations, with anti-parallel orientation. For the KR microscopy, an ac current is passed along the graphene stripe (i.e., parallel to the $b$-axis of WTe$_2$), while the induced out-of-plane spin polarisation (aligned with the $c$-axis) is locally-resolved via the \KR angle $\theta_{K}^\omega$ detected at the fundamental frequency $\omega$ of the alternating bias (Fig. \ref{fig: 1}b and Supplementary Fig. S6).


In Figure \ref{fig: 1}c, transfer curves of the graphene stripe within the above device are shown for increasing ac bias up to $V_\text{ds} = \SI{1.5}{V}$. At low ac bias amplitudes, two Dirac points are visible, revealing the presence of graphene regions with different doping levels \cite{solis2016gate}. The Dirac point at gate voltage $V_\text{g} \approx \SI{0}{V}$ can be assigned to the bare graphene leads from the metallic contacts to the junction. The second Dirac point at $V_\text{g} \approx \SI{+20}{\volt}$ indicates moderate $p$-doping and is attributable to the graphene section proximitized by WTe$_2$. With increasing ac bias amplitude, the resistance maximum associated with the Dirac point at the heterojunction progressively decreases and finally vanishes. The latter suggests a change of the current path: at large bias, the adjacent \ch{WTe2} opens a parallel current path and shunts the proximitized graphene, with a vertical current flow from graphene to the WTe$_2$ upon entering the junction region (and analogously backwards to graphene close to the opposite junction edge). Such current distribution is in agreement with the spatial maps of the \KR on the frequency $2\omega$ (Supplementary Fig. S2), which reflects the local Joule heating as a measure of the local current density \cite{stamm2017magneto}.

\begin{figure}[b]
\includegraphics[scale=0.98]{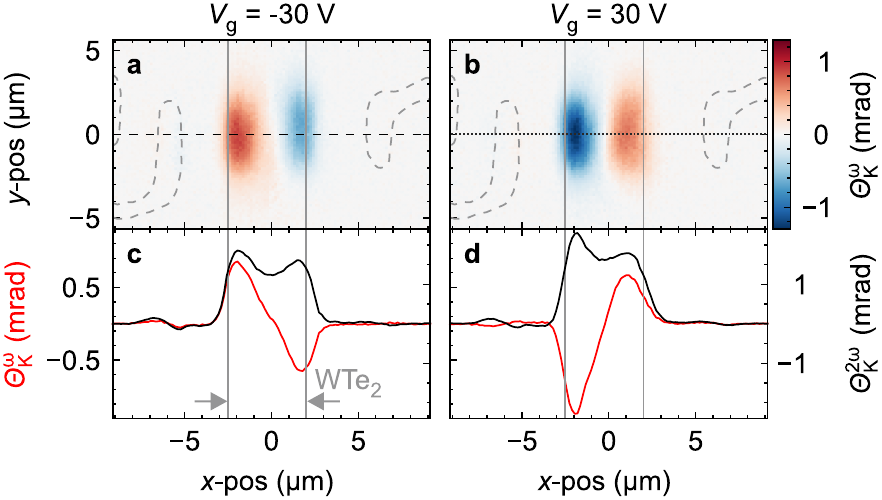}
\caption{\textbf{Gate-dependent KR microscopy}. Spatially-resolved KR, detected on the device in Fig. 1 under ac current injection along the graphene stripe ($T_\text{bath} = \SI{4.2}{K}$, $V_\text{ds} = \SI{4}{V}$) for (a) $\text{V}_\text{g} = \SI{-30}{V}$ and (b) $\text{V}_\text{g} = \SI{30}{V}$. Grey dashed lines highlight the metal electrodes. (c,d) Profiles of the Kerr angles $\theta_{K}^{\omega}$ and $\theta_{K}^{2\omega}$, along the horizontal dotted/dashed lines in (a) and (b). Vertical solid lines indicate the \ch{WTe2} ribbon.}
\label{fig: 2}
\end{figure}

\begin{figure*}[hbt]
\includegraphics[scale=1]{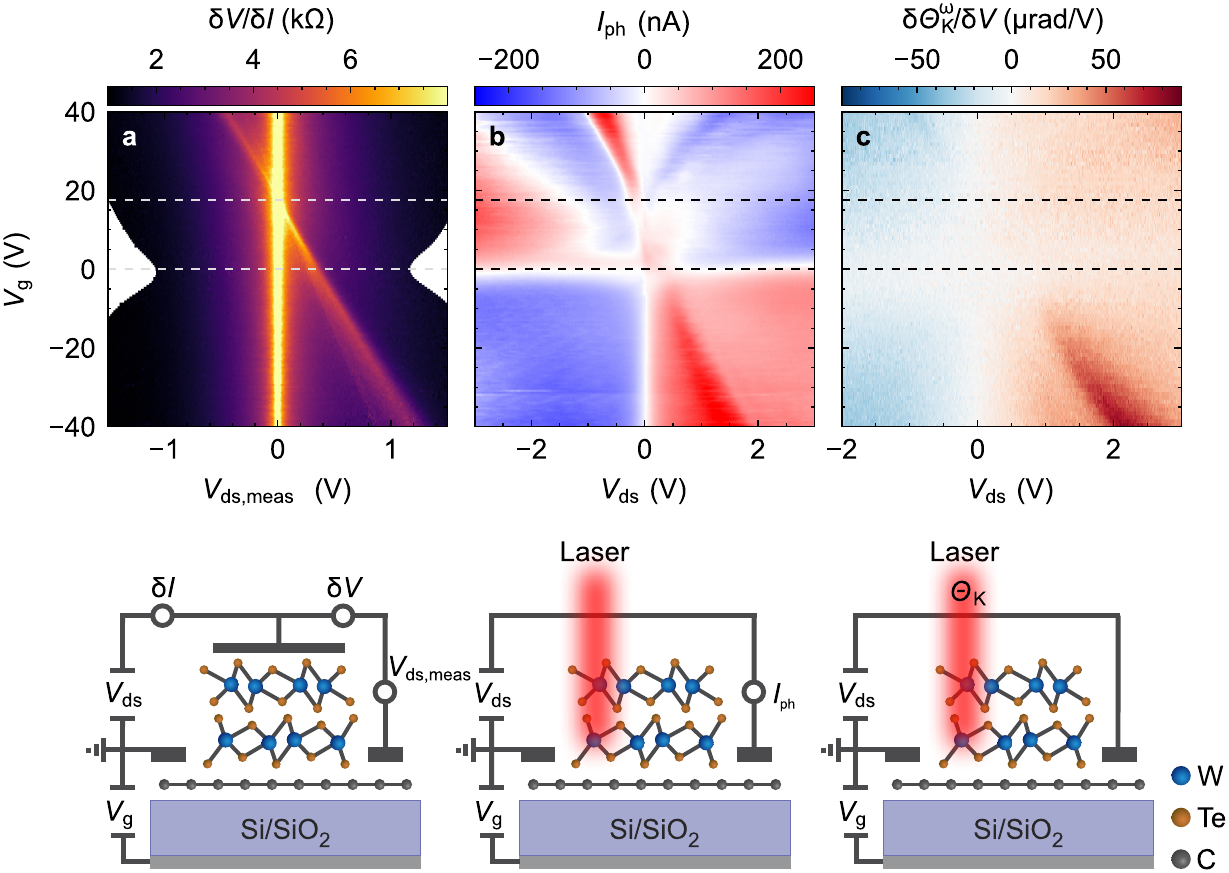}
\caption{\textbf{Electronic and optical interface spectroscopy}. (a) Differential resistance $\delta V/\delta I$ as a function of gate voltage $V_\text{g}$ and dc drain-source voltage $V_\text{ds}$. A vertical tunnel barrier is assumed due to the vdW gap between graphene and \ch{WTe2}. (b) Local photocurrent acquired at the graphene/\ch{WTe2} junction as a function of $V_\text{g}$ and $V_\text{ds}$. (c) Differential \KR signal $\theta_{K}^{\omega}$ vs. $V_\text{g}$ and dc $V_\text{ds}$. In each case, the corresponding experimental configuration is displayed at the bottom. The horizontal dotted lines indicate the positions of the two Dirac points. All data were taken at $T_\text{bath} = \SI{4.2}{K}$ on a second heterostructure device.}
\label{fig: 3}
\end{figure*}


Figure \ref{fig: 2} displays \KR maps of the above device for two gate voltage regimes. Very similar behaviour has been observed on two other devices (Supplementary Fig. S4 and S5). At $V_\text{g} = \SI{+30}{\volt}$ ($n$-type regime), strong \KR of opposite sign appears around the two edges of the junction area (Fig. \ref{fig: 2}a). Changing the gate voltage to $V_\text{g}$ = -30 V ($p$-type regime), reverses the KR polarity while its magnitude, location and spatial extent are barely affected (Fig. \ref{fig: 2}b). The location of the maximum KR coincides with the Joule heating-induced $2\omega$ signal (Supplementary Fig. S2a-d), underscoring the connection between KR signal and current flow through the heterostructure.
As elaborated by our theoretical model, the sign change of the KR signals between Fig. \ref{fig: 2}a and \ref{fig: 2}b can be traced back to the sign reversal of current injection into WTe$_2$, depending on whether the underlying graphene is $n$- or $p$-doped. Current is injected into WTe$_2$ by momentum conserving tunnelling of electrons, while microscopically, the momentum of conduction (valence) electrons in graphene is antiparallel (parallel) to the current.


The magneto-optical detection of current-induced spins in metals is challenging due to the small ratio of optically probed to spin-polarised electrons \cite{riego2016absence}, the short spin relaxation time \cite{su2017absence}, and their low optical activity compared to semiconductors. Nonetheless, it has been accomplished with the aid of current-modulation techniques \cite{puebla2017direct,stamm2017magneto}. The present \KR angles $\theta_{K}^{\omega}$ in the mrad range are approximately six orders of magnitude larger than for metal wires under ac current injection \cite{stamm2017magneto}, although the current densities are comparable (on the order of $\SI{1e7}{A cm^{-2}}$). Here, we show that the large \KR is enabled by the pronounced Berry curvature dipole of WTe$_2$ in conjunction with the significant interlayer currents. 

To shine further light on the microscopic correlation between vertical currents and  Kerr signal, we performed interlayer transport spectroscopy (Fig. 3a) of a second device (Supplementary Fig. S4). With the source-drain bias applied vertically across the junction (bottom panel of Fig. 3a), the graphene and WTe$_2$ can be regarded as planar tunnelling electrodes, possibly due to the weak coupling across the vdW gap \cite{Steinberg2015}. The bias- and gate voltage-dependent tunnelling spectroscopy reveals a prominent state that linearly shifts along the diagonal from the upper left to the bottom right. It can be assigned to the Dirac point of the proximitized graphene. The diagonal shift arises due to the interplay of the applied gate voltage, which dopes the graphene, and the bias voltage, which compensates the difference between the WTe$_2$ Fermi level and the proximitized Dirac point. 


Intriguingly, this proximitized Dirac point appears also in a nominally lateral transport configuration. Fig. \ref{fig: 3}b is a map of the local photocurrent detected at the junction (excitation wavelength = \SI{800}{nm}), while Fig. \ref{fig: 3}c shows the differential \KR signal $\delta\theta_{K}^{\omega}/\delta V_\text{ds}$. 
Akin to Fig. \ref{fig: 3}a, the photocurrent and KR signals display an onset at a bias voltage corresponding to the Dirac point of the proximitized graphene.
The linear, gate-independent background in the Kerr spectroscopy is due to Joule heating (Supplementary Fig. S3). Overall, the gate- and bias-dependencies are consistent with the theoretical picture below, which relies on minority carriers in graphene.

\begin{figure}
\includegraphics[scale=1]{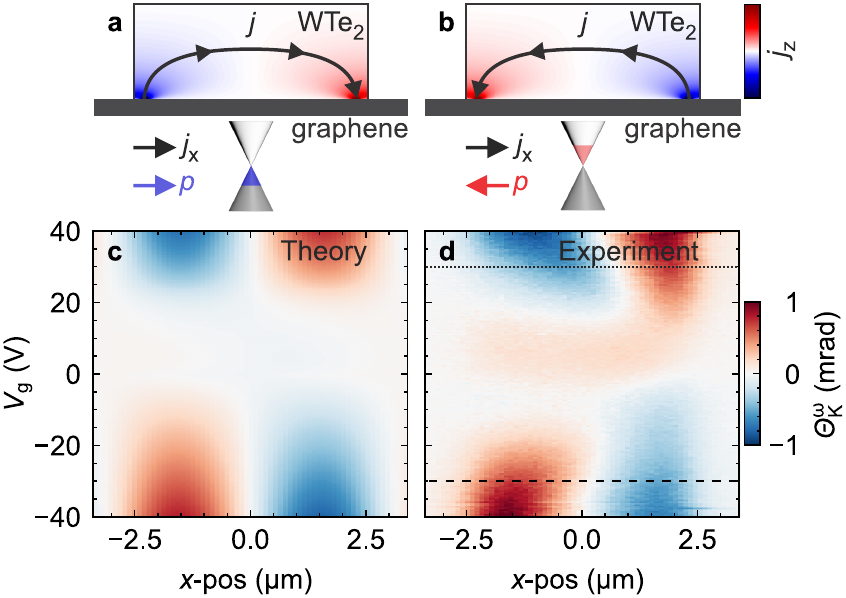}
\caption{\textbf{Theory of current-induced \KR in WTe$_2$.} (a,b) Sketch of the basic mechanism underlying the KR, with the vertical current flow profiles shown for (a) $p$-type and (b) $n$-type doping of the proximitized graphene under current flow $j_x$ in positive $x$-direction; $p$ is the carrier momentum. The colour code shows schematically the out-of-plane current density $j_z$. (c) Theoretically calculated map of gate-dependent \KR. (d) Experimentally measured \KR $\theta_{K}^{\omega}$ as a function of $x$ and $V_\mathrm{g}$ ($T_\text{bath} = \SI{4.2}{\kelvin}, V_\mathrm{ds}=\SI{3}{V}$). The dashed/dotted lines indicate where the spatial profiles in Fig. \ref{fig: 2} were taken. The data are from the device in Fig. \ref{fig: 1}.}
\label{fig: 4}
\end{figure}

We explain the above observations by a generic transport model derived from the relationship between current-induced spin-accumulation and non-linear topological transport theory in Eq.~\eqref{eq:RelSpinTopo}.
The latter provides access to both the sheet resistance and Hall conductivity, $\sigma_H = \sigma_H(\nu) = \sum_{ij} \epsilon_{ijz}\int_0^h dz \, \sigma_{ij}^{3D}(\nu)/2$ at the frequency $\nu$ of incident light with wavelength much larger than the height $h$ of the WTe$_2$ slab.
We focus on a Hall response which relies on a current-induced magnetization. Denoting the slow ac field by $E_l(\omega)$, we can write
$\sigma_{ij}(\nu)-\sigma_{ji}(\nu) = \sum_{kl} \epsilon_{ijk}\lambda_{lk}(\nu,\omega) E_l(\omega)$. The tensor $\lambda_{lk}(\nu,\omega)$ describes a non-linear (quadratic) Hall coefficient, since it relates the linear transport coefficient $\sigma_{ij}$ linearly to an external field. The Kerr effect induced by the current density $j_i$ is thus a probe of the non-linear Hall coefficient~\cite{KoenigPesin2019} 
\begin{equation}
    j_i(\nu + \omega) = \sum_{jkl} \epsilon_{ijk} \lambda_{lk}(\nu, \omega) E_j(\nu) E_l(\omega). \label{eq:lambdadef}
\end{equation}
Importantly, $\lambda_{ji}$ is proportional to the Berry curvature dipole tensor $D^{(\Omega)}_{ji} = \langle  \Omega_j v_i \rangle_{\rm FS}$, i.e.~the Fermi surface average of the Berry curvature $\Omega_j(\mathbf p)$ and velocity $v_i$. The trace of $D^{(\Omega)}_{ji}$ is topologically quantized and measures the number of Berry curvature monopoles (e.g., Dirac nodes) enclosed in a Fermi surface~\cite{SodemannFu2015}. In principle, the Kerr angle $\theta_{K} \propto \sum_l \lambda_{lz} E_l$ is a convenient experimental probe of this topological quantity and can be used to map out its fundamental symmetries. The symmetry group of T$_d$-WTe$_2$ is \textit{P$_{mn2_1}$}, and it contains a 180$^\circ$ screw rotation about the $c$-axis as well as a $b \rightarrow -b$ mirror symmetry. The latter is broken due to strain and the relative misaligment of the crystalline axes of WTe$_2$ and graphene. This imposes $\lambda_{xz} = \lambda_{yz} = 0$, while $\lambda_{zz} \neq 0$ becomes symmetry allowed. Therefore, \KR signals away from the boundaries of the heterostructure reveal the current-induced, local electric fields in $z$-direction inside the WTe$_2$.

Next, we develop a topological extension of Fick's theory of diffusion to determine the local Hall response $\sigma_H(\mathbf x) = - \int_0^h dz \varpi_z(\mathbf x)$ and thereby the local Kerr angle in the presence of an inhomogeneous current flow. The density $n(\mathbf x,t)$, the regular current density $j_i(\mathbf x,t)$ and the Berry curvature density $\varpi_i(\mathbf x,t)$ obey
\begin{subequations}
\begin{align}
j_i &= - D \partial_{x_i} n && \text{(Fick's 1$^{\rm st}$ law)},\\
\partial_t n &= D \sum_i \partial_{x_i}^2 n&& \text{(Fick's 2$^{\rm nd}$ law)}, \label{eq:Fick2}\\
\varpi_i &=  - \sum_j \tau D^{(\Omega)}_{ji} \partial_{x_j} n& & \text{(topological extension)}.
\end{align}
\end{subequations}
For simplicity, we keep the diffusion matrix diagonal and isotropic with $D_{ij} = D \delta_{ij}$.
In the presence of an electric field $E_k$, the total current contains the regular and an anomalous term, $j_{i}^{\rm (tot)} = j_i + \epsilon_{ijk} \varpi_j e E_k$, where the anomalous current describes the Berry curvature-induced anomalous velocity of Bloch electron wave packets.



Particle exchange across the heterojunction in the presence of a current $j_0$ in the graphene sheet results in current injection into the WTe$_2$ slab, which to leading order enters as a source term in Eq.~\eqref{eq:Fick2}, $D \nabla^2 n = \partial_x[\bar D \partial_x[n_e + n_h]]\delta (z)$. Here, $\bar D$ is a constant and $n_e,n_h$ are electron and hole densities in graphene. Without loss of generality, we consider contributions from momentum conserving tunnelling, such that the source term contains an imbalance current $\partial_x[n_e + n_h]$ rather than a charge current $\partial_x[n_e - n_h]$. This follows from the fact that the velocity (i.e., current) of the same momentum in $n$ and $p$-doped graphene is opposite (Fig. \ref{fig: 4}a and b). Then, it is evident that the current flows in opposite directions for $n$- and $p$-doped graphene. Furthermore, as the source term is given by the derivative of the imbalance current, it is dominated by the relaxation profile of minority carriers, which explains the gate-dependent minimal bias for the signal onset in Fig.~\ref{fig: 3}c. Fitting this model to experimental data (Supplementary Fig. S7), yields a calculated \KR response (Fig. \ref{fig: 4}c) which reproduces both the onset and the sign change of the measured \KR (Fig. \ref{fig: 4}d). We note that, in the gate voltage regime between the two Dirac points, the \KR signal, although being rather weak, has non-zero spatial average and hence cannot be interpreted based on $\varpi_z = - \tau D_{zz}^{(\Omega)} \partial_z n \propto j_z$, because the total current going into and out of the WTe$_2$ must vanish by Kirchhoff's law. We hypothesize that the weak \KR stems from a higher order non-linear current-field relation beyond Eq.~\eqref{eq:lambdadef}.

Our findings establish Kerr rotation measurements on 2D heterostructures as a direct probe of the Berry curvature dipole, which is the key quantity controlling a variety of topological response functions. Besides complementing electrical detection schemes for in-plane spins toward probing out-of-plane spins in such heterostructures, they reveal the gate-dependent spatial distribution of the current-induced spin polarisation. Considering observed spin diffusion lengths on the order of 10 µm in bare graphene \cite{avsar2020colloquium} and 2 µm in MoTe$_2$ \cite{song2020coexistence}, far-field optical probes can indeed access relevant length scales via local and non-local optoelectronic readout schemes \cite{Seifert2019, Kiemle2020b}. Furthermore, the presented generic model, merging classical diffusion theory with topology, should be applicable also to other 2D materials and their heterostructures where Berry curvature plays a crucial role, such as transition metal dichalcogenides or nodal line semimetals \cite{culcer2020transport}. We expect that microscopic modelling of the interfacial electronic structure \cite{gmitra2015graphene} will allow quantitative estimates of the Berry curvature contribution. Our work paves the way for tunable topological electronics and local control of spin polarization in 2D vdW heterostructures.


\bibliography{bibliography}

\section{Methods}

\textbf{Fabrication of graphene/\ch{WTe2} heterostructure devices}. The samples were prepared using an all-dry viscoelastic transfer method. Single layer graphene was mechanically exfoliated from natural graphite flakes (NGS Trading \& Consulting GmbH, Germany) onto a Si/\ch{SiO2} wafer. Subsequently, bulk crystals of \ch{WTe2} (hqgraphene, The Netherlands) were mechanically cleaved and exfoliated onto a PDMS stamp. Crystals with high aspect ratios, indicative of preferential cleavage along the $a$- and $b$-crystal axes, and a thickness of 15 to \SI{25}{\nano\meter} were identified by optical contrast and transferred from the stamp onto the graphene monolayer. All steps were performed under ambient.

\textbf{Magneto-optic Kerr measurements}. The magneto-optic Kerr measurements were carried out using a confocal dip-stick microscope with a sample bath-temperature of \SI{4.2}{\kelvin}. A linearly polarised cw-laser at $\lambda_\text{laser} = \SI{800}{\nano\meter}$ was focused onto the sample with a diffraction-limited spot size of $\sim\SI{800}{\nano\meter}$ and a laser power of \SI{50}{\micro\watt}. The reflected beam was guided through a 50:50 beamsplitter, a half-wave plate, a Wollaston beamsplitter, and detected by an amplified balanced photodetector. An alternating bias current with a frequency of $\omega = \SI{3.33}{\kilo\hertz}$ was applied to the sample, the polarisation change detected by the balanced photodetectors was read-out using a lock-in amplifier simultaneously at the fundamental ($\omega$) and the second harmonic ($2\omega$) frequency. Spatially-resolved scanning was performed by moving the sample using a $xy$-piezo scanner.

\textbf{Tunnelling and local photocurrent measurements}. The vertical charge transport experiments on the graphene/\ch{WTe2} heterojunction were carried out using a gate-dependent four probe differential conductivity measurement at $T_\text{bath} = \SI{4.2}{\kelvin}$ using standard lock-in detection. A dc voltage, to which a small ac amplitude (\SI{1}{mV} at \SI{77}{Hz}) was added, was applied to the graphene/\ch{WTe2} junction contacts. The resulting ac voltage was measured at the opposing graphene/\ch{WTe2} contacts. The gate voltage was applied to the Si-back gate using a source/measure unit.

\section{Acknowledgements}
Experimental work at TUM was supported by Deutsche Forschungsgemeinschaft (DFG) through the German Excellence Strategy via the Munich Center for Quantum Science and Technology (MCQST) - EXC-2111–390814868, SPP-2244 "2D Materials – Physics of van der Waals [hetero]structures" via Grant KA 5418/1-1, HO 3324/12-1, HO 3324/13-1, and the excellence cluster "e-conversion". C.K. acknowledges support through TUM International Graduate School of Science and Engineering (IGSSE). Experimental work at MPI was supported by Deutsche Forschungsgemeinschaft (DFG) through SPP-2244 "2D Materials – Physics of van der Waals [hetero]structures" via Grant BU 1125/12-1 and the DFG Grant "Weyl fermion-based spin current generation" BU 1125/11-1. We acknowledge technical support by T. Reindl, A. Güth, U. Weizmann, M. Hagel and J. Weiss from the  Nanostructuring Lab of the Max Planck Institute for Solid State Research.

\section{Author Contributions}
J.Ki. and L.P. contributed equally and are both first authors. M.B., A.W.H., and C.K. conceived and designed the experiments. L.P. fabricated the heterostructures. J.Ki. and L.P. performed the optoelectronic measurements. J.Ki., L.P., and C.K. analyzed the data. E.J.K developed the theoretical analysis and wrote the theory section with input from A.P.S. and J.K. All authors co-wrote and reviewed the manuscript.

\end{document}